\documentstyle[12pt]{report}
\title{Selection, Stability and Renormalization}
\author{Lin-Yuan Chen, Nigel Goldenfeld, Y.\ Oono, and
Glenn Paquette\thanks{Present address: Department of Physics,
	Kyoto University,
	Kyoto, 606 Japan}\\
	Department of Physics, Materials Research
Laboratory,\\ and  \\ Beckman Institute\\
	1110 W.\ Green Street \\
	University of Illinois at Urbana-Champaign \\
	Urbana, IL 61801-3080 USA\\}

\begin{document}

\maketitle

\begin{abstract}
We illustrate how to extend the concept of structural stability through
applying it to the front propagation speed selection problem.  This
consideration leads us to a renormalization group study of the problem.
The study illustrates two very general conclusions: (1) singular
perturbations in applied mathematics are best understood as
renormalized perturbation methods, and (2) amplitude equations are
renormalization group equations.

\bigskip\noindent
Pacs Numbers: 03.40.Kf, 68.10.Gw, 47.20.K

\bigskip
\rightline{P-93-10-078}

\end{abstract}

\baselineskip 4.6ex
\parskip 2.3ex

\renewcommand{\thesection}{\Roman{section}}
\renewcommand{\thesubsection}{\Roman{section}.\Alph{subsection}}
\renewcommand{\theequation}{\arabic{section}.\arabic{equation}}
\newcommand{\non}{\nonumber}
\newcommand{\sumc}[3]{\sum_{#1=#2}^{#3}}
\newcommand{\pder}[2]{\frac{\partial {#1}}{\partial {#2}}}
\newcommand{\PD}[3]{\left.\frac{\partial^{2}{#1}}{\partial{#2}^{2}}
\right|_{#3}}
\newcommand{\der}[2]{\frac{d {#1}}{d {#2}}}
\newcommand{\diff}[1]{{\rm d}{#1}}
\newcommand{\fif}{\mbox{$\Longleftrightarrow$}}
\newcommand{\hsk}{\hspace{2em}}
\newcommand{\beq}{\begin{equation}}
\newcommand{\beqa}{\begin{eqnarray}}
\newcommand{\eeq}{\end{equation}}
\newcommand{\eeqa}{\end{eqnarray}}
\newcommand{\la}{\langle}
\newcommand{\ra}{\rangle}
\newcommand{\hsp}{\hspace*{8em}}
\newcommand{\rate}[2]{I_{\tilde{#1}}^{(#2)}}
\newcommand{\lima}[1]{\lim_{#1 \rightarrow \infty} \frac{1}{#1} \ln}
\newcommand{\meas}[2]{{#1}(\mbox{d} {#2})}
\newcommand{\limin}[1]{\lim_{#1 \rightarrow \infty}}
\newcommand{\un}{\underline}
\newcommand{\noi}{\noindent}
\newcommand{\dd}{\mbox{d}}
\newcommand{\lb}{\label}
\newcommand{\fr}[1]{(\ref{#1})}
\newcommand{\bm}[1]{\mbox{\boldmath $#1$}}

\setcounter{equation}{0}
\section{Introduction}

When a very thin film made of diblock copolymers\cite{hw,leibler,ok}
in the disordered phase is quenched sufficiently, microphase separation
occurs, and segregation patterns are formed.  What happens if we cool
the film from one end? We would expect the appearance of a segregation
pattern invading the featureless disordered phase.  The quenched film
in the disordered state is thermodynamically unstable.  Thus to
facilitate the observability of such propagating front phenomena, the
growth of spontaneous fluctuations before the front must be
suppressed.  This could be accomplished, for example, by sliding a
cooling block along the film.  If we slide the block too quickly
or too slowly, however, we would not observe any intrinsic front
invasion behavior into the disordered phase; if it is too fast, the
unstable phase may spontaneously order before the front invasion, and
if it is too slow, the invasion is restricted by the presence of the
cooling front.  What is the natural speed, given the quench depth?  How
does the pattern invade with this `natural speed'?  For example,
suppose the equilibrium pattern for the low temperature state is a
triangular lattice.  When this phase invades into the quenched
disordered phase, do we observe the triangular lattice immediately, or
do we observe a lamellar phase first, which later orders into the
triangular lattice?  What are their speeds?\cite{paquette}

Now, let us examine an example.  Perhaps the simplest model of
diblock copolymer melt dynamics is the following partial differential
equation\cite{os,lg,bo,cg}:
\beq
 \partial_{t}\psi = \Delta(-\tau \psi + g \psi^{3} - D \Delta \psi) -
 B(\psi - \alpha), \lb{bcppde}
\eeq
where $\psi$ is the order parameter field, $\tau$, $g$, $D$ and $B$ are
positive constants, and $\alpha$ is a constant which could be
negative.
Figure 1 illustrates the quenching process due to the moving cooling
front simulated by the cell-dynamical system\cite{op,oscds,ocds}
corresponding to \fr{bcppde} \cite{os}.  In this particular case,
lamellae parallel to the cooling front are first formed and then break
up into a triangular pattern. In the steady state, a set of three modes,
${\cal{W}}_1 = \left\{W_{1,1}, W_{1,2} ,W_{1,3} \right\}$,
where each is parallel to one of the three edges of the triangle,
invades the disordered region. In this illustration, the
mode parallel to the cooling
front, $W_{1,1}$, invades first, followed by the remaining two.  Under
the same boundary condition, but with different polymer parameters,
sometimes a triangular lattice is formed by the invasion of the set
${\cal{W}}_2 = \left\{W_{2,1}, W_{2,2} ,W_{2,3} \right\}$, which is
rotated by 30 degrees with respect to ${\cal{W}}_1$. In general,
prior to the establishment of a steady state 3-mode invasion,
there is a competition between ${\cal{W}}_1$ and ${\cal{W}}_2$
(and any other modes which happen to be present). The time evolution
of the invasion is governed by a set of simultaneous
semilinear parabolic equations of the form\cite{paquette}
\beq
\partial_{t} \varphi_{i} = D_{i}\Delta\varphi_{i} + F_{i}(\varphi_{1},
\cdots, \varphi_{N}), \;\;\; (i=1,\cdots,N), \lb{Nmode}
\eeq
where $\varphi_{i}$ denotes the amplitude of the $i$-th mode, $N$ is
the total number of relevant modes, $D_{i}$ is the diffusion constant
for the $i$-th mode, and $F_{i}$ is the `reaction term' (a smooth
function).

In this paper,  we first wish to discuss the front selection problem
for \fr{Nmode}:  when many stable propagating fronts are allowed by the
model, what front can we actually observe under an ordinary
experimental setting?

This question is, however, only the starting point of the present
paper, whose main aim is to discuss and illustrate the fundamental role
of renormalization-group ideas in macroscopic physics.

The above question about selection has led us to the structural
stability analysis\cite{po} of \fr{Nmode} (Section III).  A
renormalized perturbation approach is given as an algorithm to check
the observability criterion due to the structural stability analysis
(Section IV).\cite{pcgoprl}  This analysis leads us to a vast frontier
of renormalization group theory (Section V).\cite{pderg}  In Section II
we give a brief review of the selection problem. The last section
contains a summary and comments.  This article contains some
pedagogical material to clearly demonstrate our points, but its main
purpose is to announce an intimate relation among structural stability,
renormalization and singular perturbation.  More accurate and detailed
statements will be published elsewhere.

\setcounter{equation}{0}
\section{Selection Problem}

The simplest case of \fr{Nmode} is obviously the following scalar
equation
\beq
\partial_{t} \varphi = \partial_{x}^{2} \varphi + F(\varphi).
\lb{fisher}
\eeq
Fisher introduced the equation with $F(\varphi) = \varphi(1-\varphi)$
(Fisher's equation).  We assume $F(0)=F(1)=0$.  If we also assume that
$F(\varphi)> 0$ $\forall \varphi \in (0,1)$, then there exists a stable
traveling wave solution interpolating between 1 and 0 with propagating
speed $c$ for all $c \in [c^*, + \infty)$.  If $F$ is differentiable at
0, then $c^* \ge \hat{c} \equiv 2\sqrt{F'(0)}$.   Thus there are
uncountably many stable propagating wave solutions for \fr{fisher}.
However, usually only one of these is reproducibly observable in actual
or computer experiments.  Thus we have the selection problem: what
stable traveling wave solution of \fr{fisher} is actually observed?

To study the selection problem, we must carefully distinguish between
the model and the system being modeled.  We use the word `system' to
denote an actual physical system on which we can perform actual
experiments.  In contrast, a model is a mathematical procedure (or
equation) describing the behavior of some observable(s) of the system
which the model is to simulate. For example, the model \fr{fisher}
simulates front propagation phenomenon such as the spreading of an
allele of a gene locus in a population (the system).  While the system
apparently exhibits reproducibly a unique propagating front, the model
allows uncountably many such fronts to exist.  What is the selection
rule for the propagating front solution which corresponds to the
actually observed front in the system?  This is the precise statement
of the selection problem.

In an actual front propagation experiment, say, fire propagation along
a fuse, we must prepare an initial condition.  Fire is set by elevating
the fuse temperature in front of the observer/experimenter.  Thus, in
practice the initial condition for the system is modified only on a
finite region of the system.  In the model, we must prepare the
corresponding initial condition to have a compact support.  Let us call
such an initial condition a {\em physical initial condition}.  We
define the `physical observability' (in the present context) of a
solution to a given model as follows.  If the traveling wave solution
is attainable as an asymptotic state of the initial value problem with
a physical initial condition, we call the traveling wave solution
{\em physically observable}.    This is sensible, since we cannot
manipulate infinite space to prepare an initial condition. We can only
modify the system just in front of us.

Aronson and Weinberger proved the following:\\ {\em Theorem} A [Aronson
and Weinberger\cite{aw}].  For \fr{fisher} if $F(0)=F(1)=0$, $F(x) > 0$
for any $x \in (0,1)$, and if $F'(0) > 0$ (these conditions will
henceforth be called the {\em AW condition}), then the boundaries of any
level set for the value in (0,1) of the solution with a physical initial
condition asymptotically travel with the speed $c^*$. $\Box$\\ This
implies that under the AW condition, the propagating speed we can
actually observe is the minimum stable speed.  We may call this the {\em
minimum speed principle}.  Empirically, this is what seems to be
generally believed.  Certainly, we do not have any counterexample for
\fr{fisher}, even without the AW condition.  We do not, however, know
any rigorous result other than this theorem.

There is a hypothesis of marginal stability due to
Langer.\cite{marginal}  The linear marginal stability analysis is
motivated by the following observation.  Suppose a small localized
perturbation is added to the $\varphi= 0$ state.  Since this state is
unstable, the disturbance grows, and consequently its fronts propagate
in both directions. We wish to observe the front from a moving frame.
If the speed of the frame is too slow, the disturbance front outruns us,
so that we observe a growing disturbance and conclude that $\varphi=0$
is unstable.  If the speed of the frame is too fast, we outrun the
disturbance, and we say $\varphi=0$ is stable.  However, the natural
front should be self-sustained; the growth of the invading disturbance
into the unstable state should be the cause of front propagation.
Hence, the speed of the front should be the one which makes the
$\varphi=0$ state marginally stable.

In the moving frame with speed $c$, \fr{fisher} reads
\beq
\partial_{t} \varphi = (\partial_{\xi}^{2} + c\partial_{\xi})\varphi +
F(\varphi),
\eeq
where $\xi = x - ct$.
We study the stability of the tip of the traveling wave in the
following form
\beq
\varphi = \epsilon(t) e^{k\xi},
\eeq
where $\epsilon$ is assumed to be very small.  We get
\beq
\epsilon'(t) = \sigma(k) \epsilon(t),
\eeq
with
\beq
\sigma(k) = k^{2} + ck + F'(0).
\eeq
The marginality condition is $Re \sigma(k) =0$ and $d\sigma(k) /d k =
0$.  From these, we conclude that $c = 2\sqrt{F'(0)}$ is the selected
speed according to the hypothesis.  Notice that this value is the lower
bound for the minimum stable speed $\hat{c}$ allowed to the model.
Mathematically, we classify \fr{fisher} into two cases\cite{stokes}:
If $c^* = \hat{c}$, the model is called a pulled case, and if $c^* >
\hat{c}$, a pushed case.  The linear marginal stability analysis works
only when the model is pulled.   There is no established method to
distinguish pulled cases from pushed cases.

\setcounter{equation}{0}
\section{Structural Stability}

To motivate our approach to the selection problem, we first wish to
reflect upon what we should mean by a good model of a natural phenomenon
(or a given system).

Suppose we repeat the same experiment many times and collect data on
the same observable for a given system.  If the observed data cluster
around some definite value, and the fluctuation around this value is
small, we may say that the observable is reproducibly observable.
Fluctuations around its most probable value are due to factors we
cannot control.  For example, they may be due to details in the initial
condition or in the system preparation  or maintenance itself.  Now,
let us assume that we have a mathematical model $M$ of the system under
study.  If this is a good model of the system, then its behavior (at
least that corresponding to the reproducible observables) must be
stable against its modification.  That is, in a certain sense, $M$ is
close to $M + \delta M$, where $\delta M$ corresponds to the details
beyond our control.

This is exactly the idea of `structural stability' of a model first
introduced in the context of dynamical systems by Andronov and
Pontrjagin.\cite{ap}  Since the coefficients of most differential
equations important in practice (in physics, biology, engineering,
etc.) cannot be determined exactly, it is crucial that their global
features be largely unaffected by tiny changes in these coefficients.
Therefore, Andronov and Pontrjagin proposed that only structurally
stable models are good models to do scientific work.  An epoch making
theorem was later proven by Peixoto\cite{peixoto}:  The set of all the
structurally stable $C^{1}$-vector fields on a $C^{\infty}$ compact
2-manifold is open and dense in the totality of $C^{1}$-vector fields.
This was a very encouraging result, suggesting that we might dismiss
all the structurally unstable models from science, as suggested by the
original proposers of the concept.\cite{open} However, soon it was
recognized that if the dimension of the manifold is larger than 2, the
structurally stable vector fields are not dense.\cite{nondense}

What does this mean to science?  It means at least:\\
(i) The World is full of systems which are in a certain respect
unstable and whose observable results are at least in part irreproducible.\\
Then, probably\\
(ii) The conventional definition of structural stability is too
restrictive for science, since the fact that many things are not
reproducible is reproducible.

If there are unstable or irreproducible aspects in the actual system
being modeled, then a good mathematical model of the system must have
features unstable with respect to the perturbation corresponding to that
causing instability in the actual system.  Thus a good model should be
structurally stable with respect to the reproducibly observable
aspects, but must be unstable with respect to the hard-to-reproduce
aspects of the actual system.

Let us consider Fisher's equation
\beq
\partial_{t} \varphi = \partial_{x}^{2}\varphi + \varphi(1-\varphi).
\lb{Fish}
\eeq
We wish to add $\delta F$ to its `reaction' term.  If $\delta F$ is
$C^{1}$-small, that is, $|\delta F|$ is small and $|\delta F'|$ is also
small in $[0,1]$, then $c^*$ changes only a little, and it is easy to
demonstrate that actually all aspects of the model are structurally
stable.  That is, all changes are continuous with respect to the
$C^{1}$-norm of $\delta F$.  Unfortunately, it is easy to demonstrate
that \fr{Fish} is not stable against certain $C^{0}$-perturbations
(i.e., without the smallness condition of $|\delta F'|$).  Consider a
small spine-like perturbation near the origin.  Its size can be made
indefinitely small while simultaneously making the slope of $\delta F$
indefinitely large.  Hence, we can indefinitely increase the slope of
the reaction term at the origin with indefinitely $C^{0}$-small
perturbations.  This implies that the lower bound $\hat{c}$ of $c^*$
can be increased without bound.  Hence, the model cannot be
structurally stable.

Is this an artifact of the mathematical model and thus a mere
pathology?  Consider the following analogy for \fr{fisher}.  We may
regard the equation to be describing the propagation of a flame along a
fuse.  In this analogy, $\varphi$ is the temperature; 0 is the flash
point of the fuse and 1 the steady burning temperature.  The reaction
term $F$ may be regarded as the generation rate of heat due to burning
(actually, it is the net rate of heat deposition on the fuse: the heat
generation due to burning minus the loss of heat to the environment.
In the steady state these must be the same, so $F(1)= 0$).  For
$\varphi \approx 0$, we may linearize \fr{fisher} as
\beq
\partial_{t} \varphi = \partial_{x}^{2} \varphi + F'(0) \varphi.
\eeq
If we put a very small amount of explosive powder along the fuse,
we can increase $F'(0)$ considerably.  The explosive burning near
temperature 0 will therefore trigger a very fast propagation of fire
along the fuse.  Thus, we can imagine an actual system in which a
drastic change of $c^*$ is possible with a very small change of $F$.
We may conclude that the structural instability of the model
\fr{fisher} is a desirable feature of a good model.  This example thus
provides an illustration of assertion (ii) above.

To relax the structural stability requirement of Andronov and
Pontrjagin, which requires every aspect \cite{dynam}of the model to be
structurally stable, we must consider two things.  First, we must
require the stability of the model only against structural
perturbations corresponding to perturbations of the actual system which
affect its reproducible observables only slightly.  We call such
perturbations {\em physically small perturbations} of the system and
the corresponding mathematical expressions {\em $p$-small
perturbations} of the model.  We require the structural stability of
the model only against $p$-small perturbations.  Secondly, we need not
require every aspect of the model to be stable against $p$-small
perturbations; we have only to require the stability of reproducibly
observable features.

Our general conjecture is: solutions structurally stable against
$p$-small perturbations describe reproducibly observable phenomena.
More precisely, we conjecture a {\em structural stability hypothesis}:
For a good model, only structurally stable consequences of the model are
reproducibly observable.  We must admit that there is potentially a
tautology here.  If we could reproducibly observe a phenomenon of a
system which is not structurally stable in the model, or if we could
not reproducibly observe something which the model says is structurally
stable, then we conclude that the model is not a faithful picture of
the system.

\setcounter{equation}{0}
\section{Structurally Stable Solutions of Semilinear Parabolic Equations}

For semilinear parabolic equations, we say a $C^{0}$-small perturbation
is $p$-small if
\beq
\sup_{u \in (0,1]} \frac{\delta F(u)}{u} < f(||\delta F||_{0}),
\eeq
where $|| \cdot ||_{0}$ is the $C^{0}$-norm, and $f$ is a continuous
function such that $f(x) \rightarrow 0$ as $x\rightarrow 0$.  Notice
that the condition has no absolute sign, and only the upper bound of
$\delta F/u$ is specified.  Thus we are not demanding the
differentiability of $\delta F$.

Now, we have the following theorem:\\
{\em Theorem} B [Paquette and Oono\cite{po}]  For \fr{fisher} with
$F(0)=F(1) =0$, let $c^*(F)$ be the minimum
traveling wave speed for the reaction term $F$.  Then, if $\delta F$
is $p$-small, $\lim_{{||\delta F ||}_{0} \rightarrow 0} c^*(F +
\delta F) =
c^*(F)$. $\Box$\\

An intuitive idea behind Theorem B is as follows.  Suppose
$\varphi(x,t) = \phi(\xi)$ (where $ \xi \equiv x - ct$) is a
traveling
wave solution to \fr{fisher}.  $\phi$ obeys
\beq
\der{^{2}\phi}{\xi^{2}} + c \der{\phi}{\xi} + F(\phi) = 0,
\eeq
or replacing $\phi$ with $q$,
\beqa
\dot{q} &=& p, \non \\
\dot{p} &=& - c p - \der{V}{q}, \lb{phase}
\eeqa
where $F(q) \equiv dV/d q$.  That is, the problem can be interpreted
as a particle of unit mass (position $q$ and velocity $p$) sliding
down a potential hill $V$ with friction constant $c$.  Hence in this
particle analogy, the speed in the original problem corresponds to the
friction constant.

A propagating front connecting 1 and 0 corresponds in the particle
analogy to an orbit connecting the saddle $S$ and the sink (at the
origin) $O$, as shown in Fig.\ 2.
If $c$ is too small, the particle overshoots 0 and goes into the region
$q <0$.  The corresponding solution of the original partial
differential equation is thus unstable in the ordinary sense of this
word.   As can be seen from Fig.\ 2, $c^*$ is the boundary between
overdamped and underdamped motion.  Now let us put a small potential
bump at the origin; this can be done with an indefinitely $C^{0}$-small
perturbation to $F$ (or indefinitely $C^{1}$-small perturbation to
$V$).  Obviously overdamped saddle-sink connection orbits no longer
exist.  That is, all the front solutions with speed faster than $c^*$
are destroyed by this perturbation.  Obviously, sufficiently
underdamped orbits still overshoot the origin, so that there must be a
boundary between over and underdamped orbits which is not far away
from the original $c^*$.  For $c <c^*$, an appropriate bump would
convert this $c$ into the critical damping factor (that is, the minimum
speed of the stable stationary front).  However, in this case we can
always choose a much smaller bump to leave $c$ as an insufficient
friction constant for the particle to stop at the origin.  Hence, the
boundary between over and underdamped cases must be infinitesimally
close to the original $c^*$, if the perturbation is infinitesimally
small.

This intuitive demonstration is technically not easy to rigorize, since
allowed perturbations are not necessarily a simple bump.  Still, it
captures the salient physics (and mathematics) behind the structural
stability of $c^*$.

If $q=0$ is not an isolated minimum of $V$, the propagating solution of
\fr{fisher} is unique.  This can easily be seen from the particle
analogy above.  Notice that it is always possible to eliminate the
isolated minimum at $q=0$  with a $p$-small perturbation.  This, together
with Theorem B, implies that $c^*$ and only $c^*$ is structurally
stable against physically benign perturbations.

In the present context, we accept that semilinear parabolic equations
are good models of front invasion into unstable states.  Then the
structural stability hypothesis implies that the physically observable
front speed is the minimum stable speed.  If the equation satisfies the
AW condition, this is true thanks to Aronson and
Weinberger's Theorem A.
But Theorem B is valid even without this condition.

For the multimode case of \fr{Nmode}, if $F_{i} = a_{i} \varphi_{i} + $
higher order terms, that is, if the $\varphi$ are linearly decoupled,
then we can prove a theorem analogous to Theorem B. \cite{po}  In this
case, however, structurally stable speeds need not be unique.
Generally speaking, there is no further principle to select one
among the structurally stable speeds.  We believe that what we can
observe in these cases depends on the initial condition.  That is, only
history can select the realized front among the structurally stable
ones.  Such examples have already been given, and in fact, the block
copolymer model is one of these.\cite{po}

We have been unable to prove the general case where no linear
decoupling assumption holds for \fr{Nmode}.  Still, we believe that what
we have seen for the decoupled case holds here too.  That is, what we can
observe are structurally stable fronts, and only history can select the
actually realized one among these.

Why does structural stability imply the  minimum speed in this case?
The key observation to explain this is that the speed $c> c^*$ is
determined by the tip, while the speed $c^*$ is determined by the bulk
of the propagating front.  The former may not be hard to understand,
because to realize a speed faster than $c^*$, we need a fine tuning of
the decay rate of the initial condition at infinity, as has been
demonstrated in the pulled case by Kolmogorov et al.\cite{kpp} and
Kametaka\cite{kametaka}. For the pushed case, see \cite{thesis}.
The assertion that $c^*$ is determined by the
bulk may sound strange in the case of a pulled front, but it is easily
seen that even in this case, $c^*$ is insensitive to  the tip.
In both the pushed and pulled cases, note
that if the initial condition is confined to a compact set, or decays
to zero more quickly than any exponential, the resulting solution
decays to zero more quickly than any exponential for all time.  Also
note that if the initial condition decays as $\sim \exp(-kx)$, where
$k$ is at least as large as $k^*$ (here $\exp(-k^*x)$ is the asymptotic
form of the steady state solution with speed $c^*$), then this
asymptotic form is maintained for all time.  In all of these cases, the
asymptotic speed is $c^*$.  The initial decay rate therefore determines
the tip shape for all time, and hence this tip shape has nothing to
do with the selected speed.  Hence, the words `pushed' and `pulled' may
both be misleading. (See \cite{po} for a more detailed explanation.)

Now it is easy to understand why the minimum speed is structurally
stable.  Since the tip is extremely fragile against small modification
of $F$ near the origin, all speeds $c >c^*$ are unstable structurally.
On the other hand, $c^{*}$ is determined by the bulk of the propagating
front, which is obviously insensitive to a small perturbation.  In
terms of the fuse analogy, imagine we put a thin film of water on the
fuse.  This would be sufficient to kill the fast propagation of fire
determined by the tip even if such propagation could be realized in the
unperturbed system.
Thus the structural instability of faster
solutions is an actual phenomenon; that is, it is not an artifact of
the modeling process.  In this sense, the reaction-diffusion equation
is a very good model of, e.g., the invasion of a stable phase into an
unstable phase.

Since unstable states are unstable against spontaneous fluctuations due
to, e.g., thermal fluctuations, it is not possible to prepare a wide
unstable phase region.  This is why the moving cooling front is used in
the diblock copolymer example at the beginning of this paper.
Therefore one might think that the nonuniqueness of the propagating
front in the model is due to an excessive idealization of the actual
system: the unstable state of the model is really a metastable state
with a very small `activation barrier'.  One might conclude that this
is the reason why in the actual system there is only one propagation
speed which we observe.    We need not deny that there are such cases,
but in many actual examples, the unstable states are really unstable
against some particular invasion mode, although they are metastable
against spontaneous fluctuations.

Consider Fisher's original example of the spreading of an allele in a
population.  Of course, the invading allele could be produced {\em de
novo} by mutation in the population, but this is extremely improbable,
so the initial population is quite stable against spontaneous
fluctuations.  If the allele is advantageous, then the initial
population is unstable against its invasion.   In the case of the fuse
analogy we have been using, the flash point $T_{F}$ is the temperature
at which the  fuel becomes unstable against the invasion of radicals,
while the ignition point $T_{I}$ is the temperature at which the fuel
can spontaneously produce radicals (reacting with oxygen).  That is,
between $T_{F}$ and
$T_{I}$, the fuse is unstable against the  invasion of fire, but
metastable (almost stable) against spontaneous thermal fluctuations.
The distinction between flash point and ignition point parallels the
distinction between the secondary and primary nucleation processes.
For example, a melt below the melting point should not be considered a
metastable state when a a crystal nucleus is already present.  The melt
is really unstable against the invasion of the crystal phase.   Thus,
the structural stability requirement cannot be regarded as simply  an
augmenting or auxiliary rule to make excessively idealized models
realistic.

\setcounter{equation}{0}
\section{Renormalization and Structural Stability}

Renormalization group (RG) methods are generally interpreted as a means
to extract structurally stable features of a model\cite{rgstr,ngtext};
the structurally stable features of the model characterize the
universality class to which it belongs.  In RG terminology Theorem B
implies that $p$-small perturbations are marginal perturbations for
$c^*$, but that some $p$-small perturbations are relevant to speeds
larger than $c^*$.  Furthermore, we know that generally speaking,
$C^{0}$-small perturbations could be relevant.

Thus Theorem B affords a method to judge whether the front with speed
$c_{0}$ is observable or not through the study of its response to
$\delta F$ corresponding to a small potential bump added to the model:
if the change of the speed
$\delta c$ vanishes in the limit of vanishing bump (that is, if $\delta
F$ is a marginal perturbation), then $c_{0}$ is observable.  Otherwise,
$c_{0}$ is not observable.   As we have found, this procedure works
numerically.  In response to a $p$-small perturbation $\delta F$, the
change in the speed of \fr{Nmode} observable in numerical computations
vanishes with $|| \delta F ||_{0}$.  Let us consider an example.

As stated above, we have been unable to prove a statement analogous to
Theorem B for multi-mode equations which display linear order
coupling.  We believe, of course,  that our structural stability
hypothesis applies to these equations as well, and in support of this
conjecture, consider one such model for the present study. We note that
similar behavior can also be easily observed for single-mode and
multi-mode, linearly decoupled equations.  Consider the following model
equation:
\beqa
\partial \psi_1/\partial t &=& 2\Delta\psi_1 + F_1(\psi_1,\psi_2)
\nonumber \\ \partial \psi_2/\partial t &=& \Delta\psi_2  +
F_2(\psi_1,\psi_2),   \label{stone}
\eeqa
where $F_1 = \psi_1 + \frac{1}{2}\psi_2 - \psi_1^3$, and $F_2 = 3\psi_2
+ \frac{1}{2}\psi_1 - \psi_1^3.  - \psi_2^3$.  We numerically studied
the behavior of \fr{stone} in response to the perturbation $F_1
\rightarrow F_1 + \delta F_1$ and $F_2 \rightarrow F_2 + \delta F_2$,
where $\delta F_i = -10 \psi_i$ if $\psi_i < \epsilon$ and 0
otherwise.  Note that $(\delta F_1, \delta F_2)$ can be considered as
the discretization of a $p$-small perturbation.
$(\delta F_1, \delta F_2)$ is analogous to the film of water
discussed in the context of the fuse analogy.
If a traveling wave
solution of (\ref{stone}) with speed $c$ is observable (structurally
stable), the speed of the observable solution of the perturbed equation
must converge to $c$ as $\epsilon \rightarrow 0$. We numerically
determined the observable propagation speed of the unperturbed
equation, as well as those of perturbed equations with several values
of $\epsilon$. The results of this study, shown in Table I, support our
structural stability hypothesis; the observable speed changes
continuously in response to a $p$-small perturbation.

We next studied a ``tip driven" solution (as opposed to the ``bulk
driven" solution considered above) of (\ref{stone}).  We were able to
produce such a solution by choosing two small positive values
$\epsilon_1$ and $\epsilon_2$, and forcing the value of $x$ at which
both $\psi_1 = \epsilon_1$ and $\psi_2 = \epsilon_2$ to move at speed
$c = 10$.  We chose $\epsilon_1 = 0.248 \times 10^{-11}$ and
$\epsilon_2 = 10^{-11}$. With these values, the eigenfunction of the
linear equation corresponding to (\ref{stone}) for the traveling wave
solution with $c = 10$ is given by:  const.$[\epsilon_1,\epsilon_2]
\exp(-kx)$, with $k = 0.323$. We then computed the speed of the
resulting front by watching the point at which $\psi_i =0.01$. Not
surprisingly, this value was 10.  However, when we applied
perturbations to the tip driven model identical to those applied to
the bulk driven model, in each case, the propagation speed computed
was also identical to that found for the bulk driven model.  For the
tip driven solution, the response of the model remains finite as the
size of the perturbation vanishes.  The above considerations thus lead
us to conclude correctly that it is unobservable.

Once more returning to the propagation of fire as a physical analogy,
this result can be interpreted as follows. For the dry fuse, we are
able to force the system to exhibit `fast' flame propagation  by
running a torch along the fuse to ignite it at the desired speed.
When we add a film of water which the torch is not able to evaporate
as it runs past, however, the behavior of this torched system
cannot be distinguished
from that of the untorched system.
Its response to this small perturbation is therefore large.

Let $\phi_{0}$ be a stable traveling wave solution of \fr{fisher} with
speed $c_{0}$.  Let us add a $p$-small structural perturbation $\delta
F$ to \fr{fisher} with $||\delta F||_{0}$ of order $\epsilon$, and assume
that in response the front solution is modified to $\phi_{0} + \delta
\phi$.  Linearizing \fr{fisher} to order $\epsilon$ in the moving frame
with velocity $c_{0}$, we obtain formally the following naive
perturbation result:
\beq
\delta\phi(\xi,t)=e^{-c_{0}\xi/2}\int_{t_0}^{t}dt'
\int_{-\infty}^{+\infty}d\xi' G(\xi,t;\xi',t') e^{c_{0}\xi'/2}\delta
F(\phi_0(\xi')).
\eeq
Here $t_{0}$ is a certain time before $\delta F(\phi_{0}(\xi))$
becomes nonzero, and $G$ is the Green's function satisfying
\beq
\frac{\partial G}{\partial t}- {\cal L}G
=\delta(t-t')\delta(\xi-\xi')
\eeq
with $G \rightarrow 0$ in $|\xi - \xi'| \rightarrow \infty$, where
\beq
{\cal L} \equiv \frac{\partial^{2}}{\partial \xi^{2}} +
F'(\phi_{0}(\xi)) - \frac{c_{0}^{2}}{4}.
\eeq

Since by $C^{0}$-infinitesimally modifying $F$, we can always cause
${\cal L}$ to have 0 as an isolated eigenvalue, we may safely disregard
all possible complications introduced by the presence of a 0 eigenvalue
which is not isolated from the essential spectrum.  Formally, $G$ reads
\beq
G(\xi,t;\xi',t') =  u_{0}(\xi)u_{0}^{*}(\xi') + \sum
e^{-\lambda_{n}(t-t')} u_{n}(\xi) u_{n}^{*}(\xi'),
\eeq
where ${\cal L} u_{0} = 0$, and ${\cal L} u_{n} = \lambda_{n} u_{n}$.
The summation symbol, which may imply appropriate integration, is over
the spectrum other than the point spectrum $\{ 0 \}$.  Since the model
is translationally symmetric, $u_{0} \propto
e^{c_{0}\xi/2}\phi_{0}'(\xi)$.  Due to the known stability of the
propagating wavefront, the operator ${\cal L}$ is dissipative, so 0 is
the least upper bound of its spectrum.  Hence, only $u_{0}$ contributes
to the secular term (the term proportional to $t- t_{0}$) in $\delta
\phi$.  Thus we can write
\beq
\delta \phi_{B} \equiv  \delta \phi = -(t- t_{0})\delta c
\phi_{0}'(\xi) + (\delta \phi)_{r},
\lb{bare}
\eeq
where the suffix $B$ means ``bare', $(\delta \phi)_{r}$ is the bounded
piece (regular part), and
\beq
\delta c = -  \lim_{\ell \rightarrow \infty}\frac{
\int_{-\ell}^{+\ell}d \xi e^{c_{0}\xi} \phi'_{0}(\xi) \delta
F(\phi_{0}(\xi) ) }{ \int_{-\ell}^{+\ell} d\xi e^{c_{0}\xi}
{\phi'}_{0}^{2}(\xi) }. \lb{dc}
\eeq
One may immediately guess that this $\delta c$ is the change in the
front speed, but the naive perturbation theory is not controlled.  A
renormalization procedure can be used to justify the
guess as follows.\cite{pcgoprl}

The first term in \fr{bare} is divergent in the $t_{0} \rightarrow
-\infty$ limit.  We introduce an arbitrary subtraction factor $\mu$ to
separate the divergence by splitting $t - t_{0}$ as $t - \mu -
(t_{0}-\mu)$, and then absorb the divergence $\mu - t_{0}$ through
renormalization of $\phi_{0}(\xi)$ to $\phi_{R}(\xi,\mu)$.  To order
$\epsilon$ we get
\beq
\phi_{B}(\xi) = \phi_{R}(\xi,\mu) - \delta c (t - \mu)
\phi'_{R}(\xi, \mu),
\eeq
where $\phi_{0}$ in the second term is replaced with $\phi_{R}$,
because $\delta c$ is already of order $\epsilon$, as seen from
\fr{dc}.  The RG equation is $\partial\phi_{B}(\xi)/\partial\mu = 0$.
Hence, to order $\epsilon$ the RG equation is, after equating $\mu$
with $t$,
\beq
\frac{\partial \phi_{R}}{\partial t} + \delta c \frac{\partial
\phi_{R}}{\partial \xi} = 0.   \lb{rgeq}
\eeq
Thus the speed of the renormalized wave is indeed $c_{0} +
\delta c$.

The formal expression \fr{dc} is legitimate only when both $\delta F$
and $-\delta F$ are $p$-small.  That is, the formula is legitimate only
when $\delta F$ is linearizable near the origin.  Since we do not know
whether the renormalized perturbation result is asymptotic or not,
strictly speaking the formal expression \fr{dc} and the true change
$\delta c \equiv c(F+\delta F) - c(F)$ itself should be distinguished.
Furthermore, the expansion is correct only if the terms obtained are
finite, so if $c$ is not structurally stable, the formal expression may
not be justified.  Still, \fr{dc} seems to give us the correct
information about the observability of $c$.

For example, if we add $\delta F = \epsilon \phi(1-\phi)$ to \fr{Fish}
with $F = \phi (1 -\phi)$, then \fr{dc} gives $c^* \simeq 2 +
\epsilon$; the exact result is, of course, $c^*= 2 \sqrt{1 +
\epsilon}$.  If we add  $\delta F=\theta (\phi-\Delta)
(\phi-\Delta)(1-\phi) - \phi(1-\phi)$, with $\Delta>0$ and $\theta$
being the unit step function, then $\delta c = \sqrt{\Delta}$ if
$c_{0}=2$, and $\delta c = \sqrt{c_{0}^{2} - 4}$ in the $\Delta
\rightarrow 0$
limit.   Hence, only when $c_{0}=2$ does $c$ change continuously
with the perturbation.

\setcounter{equation}{0}
\section{Singular Perturbation and Renormalization}

The reader may make the criticism that the renormalization approach in
the preceding section is nothing but a singular perturbation approach
(the method of stretched coordinate). Why do we need such a
(purportedly) heavy machinery as RG?  Before answering this question,
we must stress that RG is not an esoteric machinery. As mentioned in
the preceding section, it is a (the?) method to extract structurally
stable features of a given model.  For example, in the case of critical
phenomena, we wish to study global features which are insensitive to
small scale details.  That is, we are pursuing the features of the
model stable against structural perturbations corresponding to the
small scale details.

In this section, we first demonstrate that the calculation in the
preceding section is just the standard renormalization group theory
for partial differential equations.\cite{pderg,ngtext}  Then, we
demonstrate that the ordinary singular perturbation method is
understood very naturally from the RG point of view.  Actually, we wish
to claim that  singular perturbations are most naturally understood as
renormalized perturbations.

Introducing new variables $X \equiv e^{x}$ and $T \equiv e^{t}$, the
propagating front solution reads $\phi(x-ct) = \Phi(XT^{-c})$.  Thus
the front speed is interpreted as an anomalous dimension.  This is
obvious; since the variables inside logarithms must be
dimensionless, $c$ cannot be determined by dimensional analysis.  If we
introduce $T_{0}$, defined by $t_{0} = \ln T_{0}$, then $t - t_{0} =
\ln (T/T_{0})$.  From this we may interpret $T_{0}$ as an ``ultraviolet
cutoff'' scale.  Hence, the $t_{0} \rightarrow -\infty$ limit
corresponds to the {\it cutoff} $\rightarrow 0$ limit in the usual
field theoretic calculation or in our PDE calculation.  In the ordinary
multiplicative renormalization group scheme,\cite{ordrg} the
logarithmic singularity $\ln (T/T_{0})$ is absorbed into the
renormalization group constants.  Usually, we introduce an arbitrary
length scale $L$ and rewrite $T/T_{0}$ as $(T/L)( L/T_{0})$. $\ln
(L/T_{0})$ is then removed by renormalization.  Our $\mu$ above is
nothing but $\ln L$, and the splitting of the logarithmic terms should
correspond to the splitting $t - \mu + \mu - t_{0}$.  $\mu - t_{0}$
represents the divergence to be absorbed into some phenomenological
parameter.

Now, with the aid of the presumably simplest (but representative)
example, we demonstrate our point that singular perturbation is best
understood as renormalized perturbation.  Consider the following linear
ODE:
\beq
\epsilon \ddot{x} + \dot{x} + x = 0.  \lb{sex}
\eeq
We pretend that we cannot obtain its closed analytic solution and apply
a very simple-minded perturbation approach.  Expand $x$ as $ x= x_{0} +
\epsilon x_{1} + \cdots$.  We have
\beqa
\dot{x_{0}} + x_{0} &=& 0, \\
\dot{x_{1}} + x_{1} &=& -\ddot{x_{0}}.
\eeqa
Solving these equations, we can easily get the following formal
expansion:
\beq
x = A_{0} e^{-(t-t_{0})} - \epsilon A_{0} (t - t_{0}) e^{-(t-t_{0})} +
O[\epsilon^{2}], \lb{ebp}
\eeq
where $A_{0}$ is a constant determined by some initial
condition.  Now, the second term contains the prefactor $t-t_{0}$, and
is thus a secular term; the ratio of the first and the zeroth order
terms diverges in the $t_{0} \rightarrow -\infty$ limit.  As done
above, we now introduce $\mu$, split $t - t_{0} $ as $t - \mu + \mu -
t_{0}$, and absorb $\mu - t_{0}$ into $A_{0}$, which is due to the
initial condition we do not know. In this way, $A_{0}$ is renormalized
to $A$.  We rewrite \fr{ebp} as
\beq
x = A e^{-(t-\mu)} - \epsilon A (t - \mu) e^{-(t-\mu)} +
O[\epsilon^{2}].  \lb{erp}
\eeq
Since $\mu$ is not in the original problem, obviously $\partial
x/\partial\mu =0$.  This is the renormalization group equation.  After
differentiating \fr{erp} with $\mu$ and then setting $\mu$ equal to
$t$, we get
\beq
\der{A}{t} + A + \epsilon A = O[\epsilon^{2}].
\eeq
This is exactly the equation obtainable, for example, by the
reconstitution method\cite{recon}.
Solving this equation (ignoring the second order term), and putting the
result into \fr{erp} with $\mu = t$, we get
\beq
x = Be^{-(1+\epsilon)t},  \lb{str}
\eeq
where $B$ is the `phenomenological constant' we must fix appropriately
to reproduce the observable result.  Clearly \fr{str} is the formula
obtained by the usual stretched coordinate method, or a multiscale
expansion scheme.  Here the result is obtained without the introduction
of modified variables or coordinates.

One might think this agreement is only fortuitous.  To see that this is
not the case, consider \fr{ebp} again.  This formula is reliable if
$\epsilon (t - t_{0})$ is sufficiently small.  Instead of calculating
the result at $t$ at once from $t_{0}$, we could proceed step by step
just as in the Wilson renormalization group theory.\cite{wilson}  Let
us divide $t$ into $N$ time spans and first solve the problem from $0$
to $t/N$ (for simplicity, we set $t_{0}$ to be $0$).  We get
\beq
x(t/N) = Be^{-t/N}(1- \epsilon t/N) + O[(\epsilon/N)^{2}].
\eeq
Now use this as the initial value and solve $x(2t/N)$ to order
$\epsilon$, etc.  We eventually get
\beq
x(t) = B\left[e^{-t/N}(1- \epsilon t/N)\right]^{N}.
\eeq
Taking the $N \rightarrow \infty$ limit, we get \fr{str}.

To obtain a solution reliable not only for large $t$ but for all $t$,
in the standard singular perturbation procedure, the so-called inner
and outer expansion and their matching are required.\cite{singular}
Now we demonstrate that from only the inner expansion, we can construct
a uniformly valid solution by a renormalization group method.

First \fr{sex} is rewritten as
\beq
x'' + x' + \epsilon x = 0, \lb{sext}
\eeq
where $'$ implies the derivative with respect to $\tau \equiv
t/\epsilon$.  Naive perturbation gives the following result:
\beq
x= A_{0} + B_{0}e^{-\tau} - \epsilon [ A_{0}(\tau -1 +e^{-\tau}) +
B_{0}( 1- \tau e^{-\tau}- e^{-\tau})] + O[\epsilon^{2}].  \lb{1}
\eeq
Introducing $\mu$ into the secular terms through $ \tau \rightarrow
\tau - \mu + \mu$, we wish to absorb $\mu$ (here $\tau_{0}$ is set to
be 0 by an appropriate time shift) by renormalizing $A_{0}$ and $B_{0}$.
Let us proceed more systematically by introducing the multiplicative
renormalization factors, $Z_{A} = 1 + \epsilon a_{1} + \cdots$ and
$Z_{B} = 1 + \epsilon b_{1} + \cdots$, and renormalized coefficients as
$A \equiv Z_{A} A_{0}$ and $B \equiv Z_{B} B_{0}$.  Putting everything
into \fr{1}, we get to order $\epsilon$
\beqa
x &=& A(1-\epsilon a_{1}+ \cdots) + B(1-\epsilon b_{1}+
\cdots)e^{-\tau} \non \\& & - \epsilon [ A(\tau - \mu + \mu -1
+e^{-\tau}) + B( 1- (\tau-\mu+\mu) e^{-\tau}- e^{-\tau})] \non \\ & & +
O[\epsilon^{2}].
\eeqa
Thus the choice $a_{1} = \mu$ and $b_{1} = -\mu$ successfully
eliminates the secular terms, and we get the renormalized perturbation
result
\beq
x = A + Be^{-\tau} - \epsilon [ A(\tau - \mu  -1 +e^{-\tau}) + B(
1- (\tau-\mu) e^{-\tau}- e^{-\tau})] + O[\epsilon^{2}]. \lb{2}
\eeq
Notice that $A$ and $B$ are now functions of $\mu$.  Since $x$
should not depend on $\mu$, which is introduced independent of the
original problem, we have the renormalization group equation
$\partial x/\partial\mu = 0$.  From \fr{2} we get
\beq
0 = \der{A}{\mu} + \der{B}{\mu}e^{-\tau} - \epsilon [ -A + B e^{-\tau}]
+ O[\epsilon^{2}].
\eeq
Here we have used the fact that derivatives are of order $\epsilon$.
Due to
the functional independence of 1 and $e^{-\tau}$, we get
\beq
\der{A}{\mu} = -\epsilon A, \;\;\; \der{B}{\mu} = +\epsilon B.
\eeq
Solving these and equating $\mu$  and $\tau$ in \fr{2}, we get
\beq
x_{R} = Ae^{-\epsilon \tau} + Be^{-(1- \epsilon)\tau} + \epsilon (A-B)
(1- e^{-\tau}).
\eeq

Let us compare this with the result obtained by the standard
inner-outer matching method to order $\epsilon$ (that is, both the
inner and outer solutions are obtained to order $\epsilon$; notice that
this calculation is partially second order):
\beq
x = A e^{-\epsilon \tau} + B e^{- \tau} + B \epsilon \tau e^{- \tau}  +
\epsilon (A-B) (e^{-\epsilon \tau}- e^{-\tau}) -\epsilon^{2} A \tau
e^{-\epsilon \tau}.
\eeq
Except for the $\epsilon^{2}$ term, all the terms are correctly given
by the RG procedure.

\setcounter{equation}{0}
\section{Reductive Perturbation and Renormalization}

Now, let us look at \fr{rgeq}.  This is the equation of the wavefront as
seen from the moving coordinate translating with the speed of the
unperturbed front.   From this frame the motion of the perturbed front
is very slow.  Hence, \fr{rgeq} is regarded as a slow-motion equation,
like an amplitude equation obtained by the so-called reductive
perturbation methods.\cite{reductive}  That an amplitude equation is an
RG equation is not a fortuitous relation but a rule.

To see the point, let us consider the following slightly dissipative
nonlinear hyperbolic equation:
\beq
\pder{u}{t} + \lambda(u) \pder{u}{x} = \eta \pder{^{2}u}{x^{2}}, \lb{preb}
\eeq
where $\lambda(u)$ is a sufficiently smooth function of $u$, and $\eta$
is a positive constant.  We consider a small amplitude wave in the
background of the constant solution $u_{0}$,
\beq
u= u_{0} + \epsilon u_{1} + \epsilon^{2} u_{2} + \cdots,
\eeq
where $\epsilon$ denotes the amplitude of the wave.

First we study the case without dissipation ($\eta=0$).   Let us perform
a naive perturbation approach.  Let $\lambda_{0} \equiv \lambda(u_{0})$.
We
have
\beqa
\partial_{t} u_{1} + \lambda_{0} \partial_{x} u_{1} &=& 0,\\
\partial_{t} u_{2} + \lambda_{0} \partial_{x} u_{2} &=& -
\lambda'(u_{0}) u_{1} \partial_{x}u_{1},
\eeqa
and so forth.  Introducing independent variables $(\xi\equiv x -
\lambda_{0}t,t)$  to replace $(x,t)$, these equations can be rewritten
as (notice that $\partial_{t}$ now reads $ \partial_{t} + \lambda_{0}
\partial_{x}$)
\beqa
\partial_{t} u_{1} &=& 0,\\
\partial_{t} u_{2} &=& -\lambda'(u_{0}) u_{1}\partial_{\xi}u_{1}.
\lb{pert2b}
\eeqa
Thus the right hand side of the second equation is a function solely of
$\xi$ in this coordinate system, so that it gives a secular term.  Thus
to order $\epsilon^{2}$ we have the following general solution:
\beq
u = u_{0} + \epsilon F_{0}(\xi) -
\epsilon^{2}(t-t_{0})\lambda'(u_{0})F_{0}(\xi)F_{0}'(\xi). \lb{F}
\eeq
We introduce $\mu$ as we did for the propagation wave and split
$t-t_{0}$ as $t - \mu + \mu - t_{0}$.  Then we absorb $\mu - t_{0}$
into the renormalized version $F(\xi,\mu)$ of $F_{0}(\xi)$.  The
renormalized perturbation result reads to order $\epsilon^{2}$
\beq
u = u_{0} + \epsilon F(\xi,\mu) -
\epsilon^{2}(t-\mu)\lambda'(u_{0})F(\xi,\mu)\partial_{\xi}F(\xi,\mu).
\lb{rgb}
\eeq
The renormalization group equation must be $\partial u/\partial \mu
=0$, so that we get to order $\epsilon^{2}$
\beq
\partial_{\mu} F + \epsilon \lambda'(u_{0}) F \partial_{\xi}F = 0.
\lb{rgeqb}
\eeq
If we identify $\mu$ and $t$ in \fr{rgb}, we get $u = u_{0} +
F(\xi,t)$, so \fr{rgeqb} with $\mu = t$, or
\beq
\partial_{t} F + \lambda_{0} \partial_{x} F +\epsilon \lambda'(u_{0}) F
\partial_{x}F = 0,  \lb{feq}
\eeq
in the original coordinate system is the equation of motion for the
small amplitude wave.

With the introduction of a weak dissipation, the first equation of
\fr{pert2b} should not be affected (this is the precise meaning of weak
dissipation). At worst, only the second equation is modified as
\beq
\partial_{t} u_{2} = -\lambda'(u_{0}) u_{1}\partial_{\xi}u_{1} +
(\eta/\epsilon)\partial_{\xi}^{2} u_{1}.
\eeq
Thus \fr{F} is modified to be
\beq
u = u_{0} + \epsilon F_{0}(\xi) -
(t-t_{0})[\epsilon^{2}\lambda'(u_{0})F_{0}(\xi)F_{0}'(\xi) + \epsilon
\eta \partial_{\xi}^{2} F_{0}(\xi)].
\eeq
Hence instead of \fr{rgeqb}, we arrive at Burgers equation:
\beq
\partial_{t} F + \epsilon \lambda'(u_{0}) F
\partial_{\xi}F - \eta \partial_{\xi}^{2} F=0.
\eeq
This is of course a standard result obtained by a reductive
perturbation method.

\section{Summary}

At the beginning of this paper, we illustrated how to generalize the
concept of structural stability so that it is not excessively
restrictive, and we applied it to the selection problem of front
propagation speeds.  Since the basic idea of renormalization group
theory is to extract structurally stable features of a given model, this
consideration naturally led us to the RG study of front propagation.

This study in turn revealed two very general conclusions, which are
illustrated with simple examples:\\ (1) Singular perturbation methods
are best understood as renormalized perturbation methods,  and\\ (2)
Amplitude equations are just RG equations.\\ The latter in particular
strengthens our belief that RG is a prerequisite to do physics without
being affected by unknown (high-energy) details of the world.  A more
systematic presentation with numerous examples of (1) and (2) as well as
the relations to the solvability condition, center manifold theory
\cite{cmf}, etc.\ will be given elsewhere.

\smallskip
\noi
{\bf Acknowledgements}\\
We wish to dedicate this article to Kyozi Kawasaki on the occasion of
his retirement from Kyushu University.  YO is especially indebted to
him; one day in 1979, Kyozi advised YO to give up his worthless faculty
position at Kyushu University, adding that without gambling nothing
could be done.  This paper would have been nonexistent without
Kyozi's advice.   This work is, in part, supported by the National
Science Foundation grants NSF-DMR-89-20538, administered through the
University of Illinois Materials Research Laboratory, and
NSF-DMR-90-15791. GCP acknowledges support from the Japanese Society
for the Promotion of Science.

\newpage

\noindent
\begin{tabular}{||c|c||cr||}    \hline
$\delta$&$c$\\ \hline
$10^{-5}$&3.68  \\ \hline
$10^{-6}$&3.73 \\ \hline
$10^{-7}$&3.77  \\ \hline
$10^{-8}$&3.79 \\ \hline
$10^{-9}$&3.81   \\ \hline
$10^{-12}$&3.83       \\ \hline
$0$      &3.86    \\ \hline
\end{tabular}

\newpage

\noi
Table Caption\\\\\\
Table I.  The observed speed of the front as a function of the size of
perturbation.  The speed is a continuous function of perturbation.
That is, this observable speed is structurally stable.\\

\noi
Figure Captions\\\\\\
Figure 1.  A cell dynamics model simulation (for details, see
\cite{bo}) of a block copolymer film with an invading triangular
phase.  Initially, a periodic pattern is imposed on the right edge of
the system.  As time proceeds, clearly the lamellar mode parallel to
the invasion front leads the ordering process into the unstable uniform
phase.  Eventually the lamellar pattern breaks up into the final
triangular pattern (defects may be introduced in this example because
of a slight mismatching of the parameters and the systm size).  Thus,
$W_{1,1}$ invades first, followed by the remaining two modes.\\\\\\
Figure 2. An intuitive explanation of the structural stability of the
slowest stable propagation speed $c^{*}$. The trajectories
corresponding to the traveling wave solutions are illustrated for
\fr{phase}.  The left column with U is for the unperturbed model, and
the right column with P for the model perturbed with a small potential
bump at the origin.  $S$ is the saddle, and $A$ is the newly formed
stable point with the potential bump.   The friction constant $c$ (that
is, the front propagation speed in the original problem) is decreased
from A to C of the figure for both columns.  BU illustrates the
critical speed $c^*$ case; if $c$ is slightly decreased further, then
the trajectory overshoots the origin as CU.  The potential bump at the
origin prevents all the overdamped trajectories like AU from reaching
the origin, as illustrated in AP.  For most underdamped cases like CU,  a
small bump is not enough to stop overshooting.  Between AP and CP there
must be a critical friction coefficient for the perturbed model, but
it must not be far away from the unperturbed one.  Hence, $c^*$ must be
structurally stable.  Furthermore, no other $c$ can be structurally
stable.


\begin{thebibliography}{99}
\bibitem{open}  An open and dense subset of a set should not be
imagined to contain a `majority' of the points in the set.  For example,
we can easily make an open dense subset of the interval [0,1] whose
Lebesgue measure can be any small positive number, since there is a
Cantor set whose measure is indefinitely close to unity.
\bibitem{dynam}  Although we said `every aspect', this applies to
the flow structure of the dynamical system.  This does not mean that
each trajectory is stable against small perturbations.  Actually, as
seen in Anosov systems, structurally stable systems are often highly
chaotic.  However, as is explicitly noted in \cite{oota}, microscopic
instability is the basis of macroscopic stability as exhibited in the
pseudoorbit tracing properties and the stability of invariant measures
which minimize the topological pressure.
\end{thebibliography}
\enddocument